\documentstyle[12pt]{article}
\begin{document}
\begin{center} {\large\bf THE PROBLEM OF CONSTRUCTING THE CURRENT
OPERATORS IN QUANTUM FIELD THEORY} \end{center}
\vskip 1em
\begin{center} {\large Felix M. Lev} \end{center}
\vskip 1em
\begin{center} {\it Laboratory of Nuclear
Problems, Joint Institute for Nuclear Research, Dubna, Moscow region
141980 Russia (E-mail:  lev@nusun.jinr.dubna.su)} \end{center}
\vskip 1em
\begin{abstract}
Lorentz invariance of the current operators implies that they satisfy
the well-known commutation relations with the representation operators
of the Lorentz group. It is shown that if the standard construction of
the current operators in quantum field theory is used then the
commutation relations are broken by the Schwinger terms.
\vskip 0.3em

\end{abstract}
\vskip 1em

 In any relativistic quantum theory the system under consideration is
described by some unitary representation of the Poincare group. The
electromagnetic or weak current operator ${\hat J}^{\mu}(x)$
for this system (where $\mu=0,1,2,3$ and $x$ is a
point in Minkowski space) should satisfy the following necessary
conditions.

 Let ${\hat U}(a)=exp(\imath {\hat P}_{\mu}a^{\mu})$ be
the representation operator corresponding to the displacement of the origin
in spacetime translation of Minkowski
space by the four-vector $a$. Here ${\hat P}=({\hat P}^0,{\hat {\bf
P}})$ is the operator of the four-momentum, ${\hat P}^0$ is the
Hamiltonian, and ${\hat {\bf P}}$ is the operator of ordinary
momentum. Let also ${\hat U}(l)$ be the representation operator
corresponding to $l\in SL(2,C)$. Then
\begin{equation}
{\hat U}(a)^{-1}{\hat  J}^{\mu}(x){\hat U}(a)=
{\hat J}^{\mu}(x-a)
\label{1}
\end{equation}
\begin{equation}
{\hat U}(l)^{-1}{\hat J}^{\mu}(x){\hat U}(l)=L(l)^{\mu}_{\nu}
{\hat J}^{\nu}(L(l)^{-1}x)
\label{2}
\end{equation}
where $L(l)$ is the element of the Lorentz group corresponding to $l$
and a sum over repeated indices $\mu,\nu=0,1,2,3$ is assumed.

As follows from Eq. (\ref{1})
\begin{equation}
{\hat J}^{\mu}(x)=exp(\imath {\hat P}x){\hat J}^{\mu}(0)
exp(-\imath {\hat P}x)
\label{3}
\end{equation}
Therefore, if the operators ${\hat P}$ are known, the problem of
constructing ${\hat J}^{\mu}(x)$ can be reduced to that of
constructing the operator ${\hat J}^{\mu}(0)$ with the correct
properties.

 Let ${\hat M}^{\mu\nu}$ (${\hat M}^{\mu\nu}=-{\hat M}^{\nu\mu}$) be the
representation generators of the Lorentz group.  Then, as follows from
Eq. (\ref{2}), Lorentz invariance of the current operator implies
\begin{equation}
[{\hat M}^{\mu\nu},
{\hat J}^{\rho}(0)]= -\imath(g^{\mu\rho}{\hat J}^{\nu}(0)-
g^{\nu\rho}{\hat J}^{\mu}(0))
\label{4}
\end{equation}
where $g^{\mu\nu}$ is the metric tensor in Minkowski space.

 When the S-matrix is calculated in the framework of perturbation theory,
Eq. (3) is used explicitly while Eq. (4) is not. It is shown instead
(see, for example, Refs. \cite{Bjor,IZ}) that Lorentz invariance can
be ensured by using the appropriate definition of the T-product (taking
into account the Schwinger terms \cite{Schw}). However when we consider
electromagnetic or weak properties of strongly interacting systems,
perturbation theory does not apply and, for consistency,
Eqs. (1) and (2) must be satisfied. One usually believes that, although
the actual construction of the operator ${\hat J}^{\mu}(x)$ in QCD is
very difficult technical problem, in principle QCD makes it possible to
construct this operator in such a way that Eqs. (1) and (2) will be
satisfied. The purpose of the present paper is to pay attention to the
fact that the standard procedure of constructing the current operators in
quantum field theory leads to the operators which do not satisfy Eq. (2).

 In the standard formulation of quantum field theory the operators
${\hat P}_{\mu},{\hat M}_{\mu\nu}$ are given by
\begin{equation}
{\hat P}_{\mu}=\int\nolimits {\hat T}_{\mu}^{\nu}d\sigma_{\nu}(x),\quad
{\hat M}_{\mu\nu}=\int\nolimits {\hat M}_{\mu\nu}^{\rho}d\sigma_{\rho}(x)
\label{5}
\end{equation}
where ${\hat T}_{\mu}^{\nu}$ and ${\hat M}_{\mu\nu}^{\rho}$ are the
energy-momentum and angular momentum tensors and
$d\sigma_{\mu}(x)=\lambda_{\mu}\delta(\lambda x-\tau)d^4x$ is
the volume element of the space-like hypersurface defined by the time-like
vector $\lambda \quad (\lambda^2=1)$ and the evolution parameter $\tau$.
The initial commutation relations for the field operators are given on
the hypersurface $\sigma_{\mu}(x)$.

\begin{sloppypar}
For simplicity we will consider the well-known case of QED where
the interaction Lagrangian is given by
$L_{int}(x)=-e{\hat J}^{\mu}(x){\hat A}_{\mu}(x)$, $e$ is the electron
charge and ${\hat A}_{\mu}(x)$ is the operator of the photon field.
 Then it is easy to show (see, for example, Refs. \cite{AB,Bog}) that
\begin{eqnarray}
&&{\hat P}^{\mu}=P^{\mu}-\lambda^{\mu}\int\nolimits L_{int}(x)
\delta(\lambda x-\tau)d^4x,\nonumber\\
&&{\hat M}^{\mu\nu}=M^{\mu\nu}-\int\nolimits L_{int}(x)
(x^{\nu}\lambda^{\mu}-x^{\mu}\lambda^{\nu})
\delta(\lambda x-\tau)d^4x
\label{6}
\end{eqnarray}
where we use $P^{\mu}$ and $M^{\mu\nu}$ to
denote the four-momentum operator and the generators of the Lorentz
group in the case when all interactions are turned off.
\end{sloppypar}

 It is well-known (see, for example, Refs. \cite{AB,BD}) that the
operator ${\hat J}^{\mu}(x)$ in QED is given by
\begin{equation}
{\hat J}^{\mu}(x)=\frac{1}{2}[{\hat {\bar \psi}}(x),\gamma^{\mu}
{\hat \psi}(x)]
\label{7}
\end{equation}
where ${\hat \psi}(x)$ is the Heisenberg operator of the electron-positron
field. If $x=0$ this operator coincides with the free operator
$\psi (0)$ in the interaction representation and therefore, as follows
from Eq. (7), ${\hat J}^{\mu}(0)=J^{\mu}(0)$ where $J^{\mu}(x)$ is the
current operator in the free theory.

 The most often considered case is $\tau =0$, $\lambda =(1,0,0,0)$. Then
$\delta(\lambda x-\tau)d^4x=d^3{\bf x}$ and the integration in Eq. (6) is
taken over the hyperplane $x^0=0$. Therefore, as follows from Eq. (6),
${\hat {\bf P}}={\bf P}$, and, as follows from Eq. (3), the operator
${\hat J}^{\mu}(0,{\bf x})\equiv {\hat J}^{\mu}({\bf x})$ is free, i.e.
${\hat J}^{\mu}({\bf x})=J^{\mu}({\bf x})$. It is also obvious that
${\hat A}^{\mu}({\bf x})=A^{\mu}({\bf x})$.

Since $J^{\mu}(0)$ satisfies Eq.
(4) if ${\hat M}^{\mu\nu}=M^{\mu\nu}$, it follows from the second
expression in Eq. (6) that ${\hat J}^{\nu}(0)$ will satisfy Eq. (4) if
\begin{equation}
\int\nolimits x^i A_{\mu}({\bf x})[J^{\mu}({\bf x}),J^{\nu}(0)]
d^3{\bf x}=0 \quad (i=1,2,3)
\label{8}
\end{equation}
for all $\nu$.

 It is well-known that if the standard equal-time commutation relations
are used naively then the commutator in Eq. (8) vanishes for all $\mu,\nu$
and therefore this equation is satisfied. However the famous Schwinger
result \cite{Schw} is
\begin{equation}
[J^i({\bf x}),J^0(0)]=C\frac{\partial}{\partial x^i}\delta({\bf x})
\label{9}
\end{equation}
where $C$ is some (infinite) constant. Therefore Eq. (8) is not satisfied
and the current operator ${\hat J}^{\mu}(x)$ does not satisfy Lorentz
invariance.

 Let us now consider the following question. While the arguments given in
Ref. \cite{Schw} prove that the commutator in Eq. (9) cannot vanish, one
might doubt whether the singularity of the commutator is indeed given by
the right hand side of this expression. However it is easy to show that
only this form of the commutator is compatible with the continuity equation
$\partial {\hat J}^{\mu}(x)/\partial x^{\mu}=0$. Indeed, as follows from
this equation and Eq. (3), $[{\hat  J}^{\mu}(0),{\hat P}_{\mu}]=0$. Since
${\hat J}^{\mu}(0)=J^{\mu}(0)$ and $J^{\mu}(0)$ satisfies the condition
$[J^{\mu}(0),P_{\mu}]=0$, it follows from Eq. (6) that the continuity
equation is satisfied only if
\begin{equation}
\int\nolimits A_{\mu}({\bf x})[J^{\mu}({\bf x}),J^0(0)]
d^3{\bf x}=0
\label{10}
\end{equation}
In turn this relation is satisfied only if $[J^0({\bf x}),J^0(0)]=0$ and
Eq. (9) is valid, since $div({\bf A}({\bf x}))=0$.

 As pointed out by Dirac \cite{Dir}, any physical system can be
described in different forms of relativistic dynamics. By definition,
the description in the point form implies that the operators
${\hat U}(l)$ are the same as for noninteracting particles, i.e.
${\hat U}(l)=U(l)$ and ${\hat M}^{\mu\nu}=M^{\mu\nu}$, and thus
interaction terms can be present only in the four-momentum operators
${\hat P}$ (i.e. in the general case ${\hat P}^{\mu}\neq P^{\mu}$ for
all $\mu$). The description in the instant form implies that the
operators of ordinary momentum and angular momentum do not depend on
interactions, i.e. ${\hat {\bf P}}={\bf P}$, ${\hat {\bf M}}={\bf M}$
$({\hat {\bf M}}=({\hat M}^{23},{\hat M}^{31},{\hat M}^{12}))$, and
therefore interaction terms may be present only in ${\hat P}^0$ and the
generators of the Lorentz boosts ${\hat {\bf N}}=({\hat M}^{01},
{\hat M}^{02},{\hat M}^{03})$. In the front form with the marked $z$
axis we introduce the + and - components of the four-vectors as $x^+=
(x^0+x^z)/\sqrt{2}$, $x^-=(x^0-x^z)/\sqrt{2}$. Then we require that
the operators ${\hat P}^+,{\hat P}^j,{\hat M}^{12},{\hat M}^{+-},
{\hat M}^{+j}$ $(j=1,2)$ are the same as the corresponding free
operators, and therefore interaction terms may be present only in the
operators ${\hat M}^{-j}$ and ${\hat P}^-$.

 In quantum field theory the form of dynamics depends on the choice of
the hypersurface $\sigma_{\mu}(x)$. In particular, it is clear from the
above consideration that the choice $\tau =0$, $\lambda =(1,0,0,0)$
leads to the instant form \cite{Dir}. The front form can be formally
obtained from Eq. (6) as follows. Consider the vector $\lambda$ with the
components
\begin{equation}
\lambda^0=\frac{1}{(1-v^2)^{1/2}},\quad \lambda^j=0,\quad
\lambda^3=-\frac{v}{(1-v^2)^{1/2}}\quad (j=1,2)
\label{11}
\end{equation}
Then taking the limit $v\rightarrow 1$ in Eq. (6) we get
\begin{eqnarray}
&&{\hat P}^{\mu}=P^{\mu}-\omega^{\mu}\int\nolimits L_{int}(x)
\delta(x^+)d^4x,\nonumber\\
&&{\hat M}^{\mu\nu}=M^{\mu\nu}-\int\nolimits L_{int}(x)
(x^{\nu}\omega^{\mu}-x^{\mu}\omega^{\nu})
\delta(x^+)d^4x
\label{12}
\end{eqnarray}
where the vector $\omega$ has the components $\omega^-=1$,
$\omega^+=\omega^j=0$. It is obvious that the generators (12) are given
in the front form and that's why Dirac \cite{Dir} related this form to
the choice of the light front $x^+=0$.

 By analogy with the above consideration it is easy to show that the
standard light front quantization also leads to the current operator which
does not satisfy Eq. (2). Let us also note that if the theory should be
invariant under the space reflection or time reversal, the
corresponding representation operators in the front form ${\hat U}_P$ and
${\hat U}_T$ are necessarily interaction dependent (this is clear, for
example, from the relations ${\hat U}_PP^+{\hat U}_P^{-1}$ =
${\hat U}_TP^+{\hat U}_T^{-1}$ = ${\hat P}^-$)
and the operator ${\hat J}^{\mu}(0)$ should satisfy the conditions
\begin{equation}
{\hat U}_P({\hat J}^0(0),{\hat {\bf J}}(0)){\hat U}_P^{-1}=
{\hat U}_T({\hat J}^0(0),{\hat {\bf J}}(0)){\hat U}_T^{-1}=
({\hat J}^0(0),-{\hat {\bf J}}(0))
\label{13}
\end{equation}
Therefore it is not clear whether these conditions are compatible with
the relation ${\hat J}^{\mu}(0)=J^{\mu}(0)$. However this difficulty is
a consequence of the difficulty with Eq. (2) since, as noted by Coester
\cite{Coes}, the interaction dependence of the
operators ${\hat U}_P$ and ${\hat U}_T$ in the front form does not mean
that there are discrete dynamical
symmetries in addition to the rotations about transverse axes.
Indeed, the discrete transformation $P_2$ such that
$P_2\, x:= \{x^0,x_1,-x_2,x_3\}$ leaves the light front $x^+=0$ invariant.
The full space reflection $P$ is the product of $P_2$ and a rotation about
the 2-axis by $\pi$. Thus it is not an independent dynamical transformation
in addition to the rotations about transverse axes.
Similarly the transformation $TP$ leaves $x^+=0$ invariant and
$T=(TP)P_2R_2(\pi)$.

 The fact that the choice ${\hat J}^{\mu}(0)=J^{\mu}(0)$ is incompatible
with Lorentz invariance was pointed out by several authors investigating
relativistic effects in nuclear physics (see, for example, Refs.
\cite{GK,CCKP}). In our opinion this problem is in fact algebraic. Indeed,
in quantum field theory the standard quantization procedure implies that
${\hat J}^{\mu}(0)=J^{\mu}(0)$ while some of the operators
${\hat M}^{\mu\nu}$ contain interaction terms. Since there is no reason to
believe that they commute with $J^{\mu}(0)$, it is reasonable to conclude
that Eq. (4) is not satisfied and therefore Eq. (2) is not satisfied too.

 At the same time, if ${\hat J}^{\mu}(0)=J^{\mu}(0)$ then
Eq. (4) is obviously satisfied in the point form. In Ref.
\cite{Dir} the point form was related to the hypersurface
$t^2-{\bf x}^2>0,\,t>0$, but as argued by Sokolov \cite{Sok} the point form
should be related to the hyperplane orthogonal to the four-velocity of the
system under consideration. In view of the above discussion this problem
deserves investigation.

 Let us summarize our discussion.
Lorentz invariance of the operator ${\hat J}^{\mu}(x)$ implies that the
commutation relations (2) are satisfied. However if the standard
construction of ${\hat J}^{\mu}(x)$ in quantum field theory is used then the
inevitable presence of the Schwinger terms leads to the conclusion that
these relations are not satisfied.

\begin{center} {\bf Acknowledgments} \end{center}

 The author is grateful to F.Coester, S.B.Gerasimov and O.Yu.Shevchenko
for valuable discussions.  This work was supported by grant No.
93-02-3754 from the Russian Foundation for Fundamental Research.

\end{document}